\documentclass[prd,twocolumn,showpacs,preprintnumbers,amsmath,amssymb,floatfix,nofootinbib]{revtex4}

\usepackage{subfigure,graphicx,epsfig,amsmath,amsfonts,amssymb,xcolor,slashed,ulem}

\usepackage{bm}
\usepackage{graphicx}
\usepackage{amsmath}
\usepackage{amsfonts}
\usepackage{amssymb}
\usepackage{color}
\usepackage{ulem}

\usepackage[section]{placeins}

\usepackage{booktabs}
\usepackage[colorlinks, citecolor=blue,anchorcolor=red,menucolor=red, linkcolor=red,filecolor=red,runcolor=red,urlcolor=blue,frenchlinks=red]{hyperref}

\newcommand{\be}{\begin{equation}}
\newcommand{\ee}{\end{equation}}
\newcommand{\ba}{\begin{eqnarray}}
\newcommand{\ea}{\end{eqnarray}}
\newcommand{\nn}{\nonumber}

\newcommand{\mev}{\textrm{ MeV}}
\newcommand{\gev}{\textrm{ GeV}}

\begin{document}
\bibliographystyle{unsrt}
\arraycolsep1.5pt

\title{Scalar resonances in the $D^+\to K^-K^+K^+$ decay}

\author{L. Roca }
\email{luisroca@um.es}
\affiliation{Departamento de F\'i谋sica, Universidad de
Murcia, E-30071 Murcia, Spain}

\author{E. Oset}
\email{oset@ific.uv.es}
\affiliation{Departamento de F\'{\i}sica Te\'orica and IFIC,
Centro Mixto Universidad de Valencia-CSIC Institutos de Investigaci\'on de Pate\
rna, Aptdo.22085, 46071 Valencia, Spain}

\begin{abstract}
We study theoretically the resonant structure of the double Cabibbo suppressed $D^+\to K^- K^+K^+$ decay. We start from an elementary production diagram, considered subleading in previous approaches, which cannot produce a final $K^- K^+$ pair at the tree level but which we show to be able to provide the strength of the decay through final meson-meson state interaction. The different meson-meson elementary productions are related through SU(3)   and the final rescattering is implemented from a suitable implementation of unitary extensions of ChPT which generate dynamically the scalar resonances $f_0(980)$ and $a_0(980)$. We obtain a good agreement with recent experimental data from the LHCb collaboration with a minimal freedom in the fit and show the dominance of the $a_0(980)$ contribution close to the threshold of the $K^- K^+$  spectrum.

\end{abstract}

\maketitle

\section{Introduction}

  Weak decay of heavy mesons into hadrons has become an important source of information on the hadron-hadron interaction. In particular the decay of $D$ mesons into three pseudoscalars has drawn the attention of different groups with the aim of learning about the meson-meson interaction \cite{patricia,robilotta,escribano,diakono,robert,dai}. These works deal about the $D \to K \pi \pi$ decay, which is Cabibbo favored and is mainly used to learn about the $\pi \pi$ interaction. Other $D$ decays are studied in \cite{dias}, using the $D \to \pi^+ \pi^+ \pi^-$ and  $D \to \pi^+ K^+ K^-$ reactions to learn about the $\pi \pi$ and $K \bar K $ interaction, in \cite{raquel,Hsiao:2019ait} interpreting the $D_s^+ \to \pi^+ \pi^0 \eta$ decay, in \cite{enwang} studying the single Cabibbo suppressed $D^+ \to \pi^+ \pi^0 \eta$ decay, or in \cite{sakai} studying the $D_s^+ \to \pi^+ \pi^0 a_0(980) (f_0(980)$ reactions. The $D^0 \to K^- \pi^+ \eta$ reaction is also studied in \cite{Toledo:2020zxj} from where information on the $a_0(980)$ and $\kappa (K_0^*(700))$ is obtained.

  The reaction that we study here is $D^+ \to K^-K^+K^+$, which is doubly Cabibbo suppressed, but which can teach us much about the $K \bar K$ interaction, one of the pseudoscalar interaction channels most poorly known. The reaction has been studied by the LHCb collaboration in \cite{Aaij:2019lwx} and analyzed using two methods, the standard one, the isobar model, and then the triple-M model developed in \cite{Aoude:2018zty}. The isobar model is the standard method used in the LHCb analysis and in most of the experimental collaborations. The full decay amplitude is written in terms of the only two independent variables 
\begin{equation}  
  T(s_{12},s_{13})= c_{NR} + \sum_k c_k T_k(s_{12},s_{13}) ,
\end{equation}
where $c_{NR}$ is a non resonant background term and $T_k$ are intermediate resonant amplitudes properly parametrized. The parameters in the different terms and the complex weights $c_k$ are obtained by performing a best fit to the Dalitz plot data. The method is efficient to extract information on the role played by different resonances, but has its limitations. Using words of Ref. \cite{Aoude:2018zty} "This
approach, albeit largely employed \cite{moreex}, has conceptual
limitations. The outcome of isobar model analyses are
resonance parameters such as fit fractions, masses and
widths, which are neither directly related to any underlying
dynamical theory nor provide clues to the identification of
two-body substructures. Thus, the systematic interpretation
of the isobar model results is rather difficult."

Steps to make different analyses of the data to allow a better matching with theoretical tools used in the study of meson interactions have been done in \cite{boito}, and tools to use three body dynamics have also been used in the $D \to K \pi \pi$ reactions \cite{patricia,robilotta,frederico,kubis,Nakamura:2015qga}. Yet, the majority of analyses rely upon the consideration of two body amplitudes having one of the mesons as spectator and this is also used in \cite{Aoude:2018zty}. The fact that the three body amplitudes can be constructed from on shell two body amplitudes, since off shell parts are shown to cancel with contact terms present in the theory, makes this approach more realistic \cite{alberto,reviewthree}. However, there are other reasons to neglect terms involving explicitly three particles interacting because after the interaction of a pair of mesons in regions where a resonance appears to be important, the resulting invariant mass of one of the particles of the pair with the third one does not have a given value but usually spans a large region of invariant mass, thus diluting the possible contribution of another resonance, which, however, is taken into consideration with the direct interaction of this original pair considering the third particle as spectator. 

   The work of \cite{Aoude:2018zty} uses effective Lagrangians to deal with the weak and strong interaction. For the weak interaction the starting point is the diagram of Fig.~\ref{fig:Tquarks}(b), which involves quark pair annihilation (W-annihilation). The diagram of Fig.~\ref{fig:Tquarks}(a), which involves external emission, is considered as a possible mechanism, but it does no provide $K^-K^+K^+$ upon hadronization of the quarks, and final state interaction is needed to produce this state. For this reason it is neglected in the analysis of \cite{Aoude:2018zty} and the annihilation mechanism of Fig. 1(a) is used as the starting point. The mesons stemming from the double hadronization of this mechanism are allowed to follow final state interaction. The final state interaction of the mesons in \cite{Aoude:2018zty} is done using Lagrangians of \cite{ecker} in which chiral perturbation is used including resonances explicitly. Yet, extra unitarization is considered in \cite{Aoude:2018zty}, using techniques of the chiral unitary approach of \cite{Oller:1998zr,Oller:1998hw,kanchan}, which justifies that the parameters that they get from a fit to the data are not the parameters used in \cite{ecker}.
The unitarization is however done with one approximation with respect to the former works, using the K-matrix approach in which the real part of the loop functions is neglected and only the imaginary part is kept. The use of the K-matrix approach has one intrinsic problem, which is that ignoring the real part of the loops prevents that resonances are dynamically generated. This is why they have to be introduced by hand, and the approach cannot tell us about the nature of these resonances. However, resonances like the $f_0(980)$ and $a_0(980)$ appear as a consequence of the meson-meson interaction in the chiral unitary approach of \cite{Oller:1998zr,Oller:1998hw,kanchan} and the consistency of the approach with data gives support to this picture. We should mention that in a recent work \cite{Magalhaes:2020tmw}, the authors or Ref.~\cite{Aoude:2018zty} are already considering the real parts of the loops. 

 In the present approach to the $D^+\to K^-K^+K^+$ reaction, we only have three parameters to fit to the data: the global strength, the strength of the $\phi K^+$ production amplitude and a relative phase of the s-wave  to p-wave mechanisms. Since the  global strength is irrelevant when we compare to events in the data, and the global strength of the $\phi K^+$ is easily determined from the clean peak in the $K^+ K^-$ mass distribution, our approach has basically one degree of freedom to fit all the data. This contrasts with the 10 free parameters that one has in the approach of \cite{Aoude:2018zty}. The isobar model has even more parameters. 

    We have here two of the most important differences between the work of \cite{Aoude:2018zty} and the one we present here. The starting point for us is not the diagram of Fig.~\ref{fig:Tquarks}(b) but the one of Fig.~\ref{fig:Tquarks}(a).  The reason is the following: In the classification of weak decay topologies of \cite{chau,Morrison:1989xq} the order of importance is: external emission, internal emission, W-exchange and annihilation. Given the fact that the $K \bar K$ interaction al low energies is driven by the $f_0(980)$ and $a_0(980)$ resonances, the final state interaction is very important and does not destroy the order of the primary decays. In the analysis of the BESIII data for $D_s^+ \to \pi^+ \pi^0 \eta$, \cite{Ablikim:2019pit}, the process removing the $\rho^+ \eta$ channel, was supposed to proceed via W-annihilation with a rate of an order of magnitude bigger than for usual W-annihilation processes. Yet, the decay was studied in \cite{raquel} with an approach similar to the one we follow here and it was shown that the process proceeded via internal emission and final state interaction. 

  Our approach to these weak processes consist in a first identification of the dominant mechanisms at the quark level, then we proceed with hadronization to produce the mesons that appear in a first step, and then consider the final state interaction of these mesons to produce at the end the desired final state. We must take into account that the hadronization, including new $q \bar q$ pairs to produce mesons, results in a reduction factor in the decay amplitude. Then, in the mechanism of Fig.~\ref{fig:Tquarks}(b) used in \cite{Aoude:2018zty} one has two hadronizations, while in the mechanism that we have of Fig.~\ref{fig:Tquarks}(a) there is only one hadronization. All these reasons justify that we neglect the mechanism of Fig.~\ref{fig:Tquarks}(b) and start the process with the one of Fig.~\ref{fig:Tquarks}(a). The choice of the starting mechanism is not innocuous: different initial meson channels are produced, which upon final state interaction give rise to the three kaons. This means that coupled channels have to be used in the approach, as also done in \cite{Aoude:2018zty}, but the amount $f_0(980)$ or $a_0(980)$  production, for instance, depends on the mechanisms assumed and the weights by which the different meson channels appear in the hadronization. 

  As we mentioned, in the experiment of Ref. \cite{Aaij:2019lwx}, two methods of analysis were made, the first one using the isobar model and the other one using the formalism of \cite{Aoude:2018zty}. The problems using the isobar model which were exposed in \cite{Aoude:2018zty} are further evidenced in the study done in \cite{Aaij:2019lwx}. Indeed, when using the isobar model three options, A, B, C are used.  In model A they include the $\phi K^+$, $f_0(980) K^+$ and $f_0(1370) K^+$ channels. In model B they add a nonresonant amplitude. In model C they replace the $f_0(1370) K^+$ channel by the  $a_0(980) K^+$ one. They note that both models B and C are equally acceptable. This means that this type of analysis cannot tell us about the relevance of the  $a_0(980)$ resonance in this reaction. By the contrary, the analysis done there using the model of \cite{Aoude:2018zty} shows a dominant role played by the $a_0(980)$ resonance. Incidentally, we should also mention that by using fully unitarized amplitudes in coupled channels in the approach that we follow, we do not have to worry about background. The amplitudes contain the resonance pole but provide at the same time background terms away from the resonance peaks. 

   The strategy of our work, which is widely used (see Ref. \cite{review} for a review on this issue) is to establish the dominant mechanisms at the quark level that produce the desired number of mesons after hadronization including $q \bar q$ pairs with the quantum numbers of the vacuum. Then all pairs of mesons are allowed to undergo final state interaction keeping a third particle as spectator. The amplitudes are properly symmetrized to account for the identity of the particles. The final sate interaction requires to use the meson-meson scattering amplitudes, which we take from the prior study with the chiral unitary approach. This method has a minimum input, basically a global strength and some relative strength from the s-wave to p-wave amplitudes. An agreement with data with this minimum input is considered as giving support for the chiral unitary approach to the meson-meson interaction and in particular for the nature of some of the resonances that it provides as dynamically generated from the meson interaction rather than genuine states of particular quark configurations.  The good agreement with data that we will show in the present work will then give support to this kind of molecular picture for the $f_0(980)$ and $a_0(980)$ resonances, which adds to the support from many other reactions where these states are produced \cite{review}.

\section{Formalism}

\begin{figure}[h]
     \centering
     \subfigure[]{\includegraphics[width=.85\linewidth]{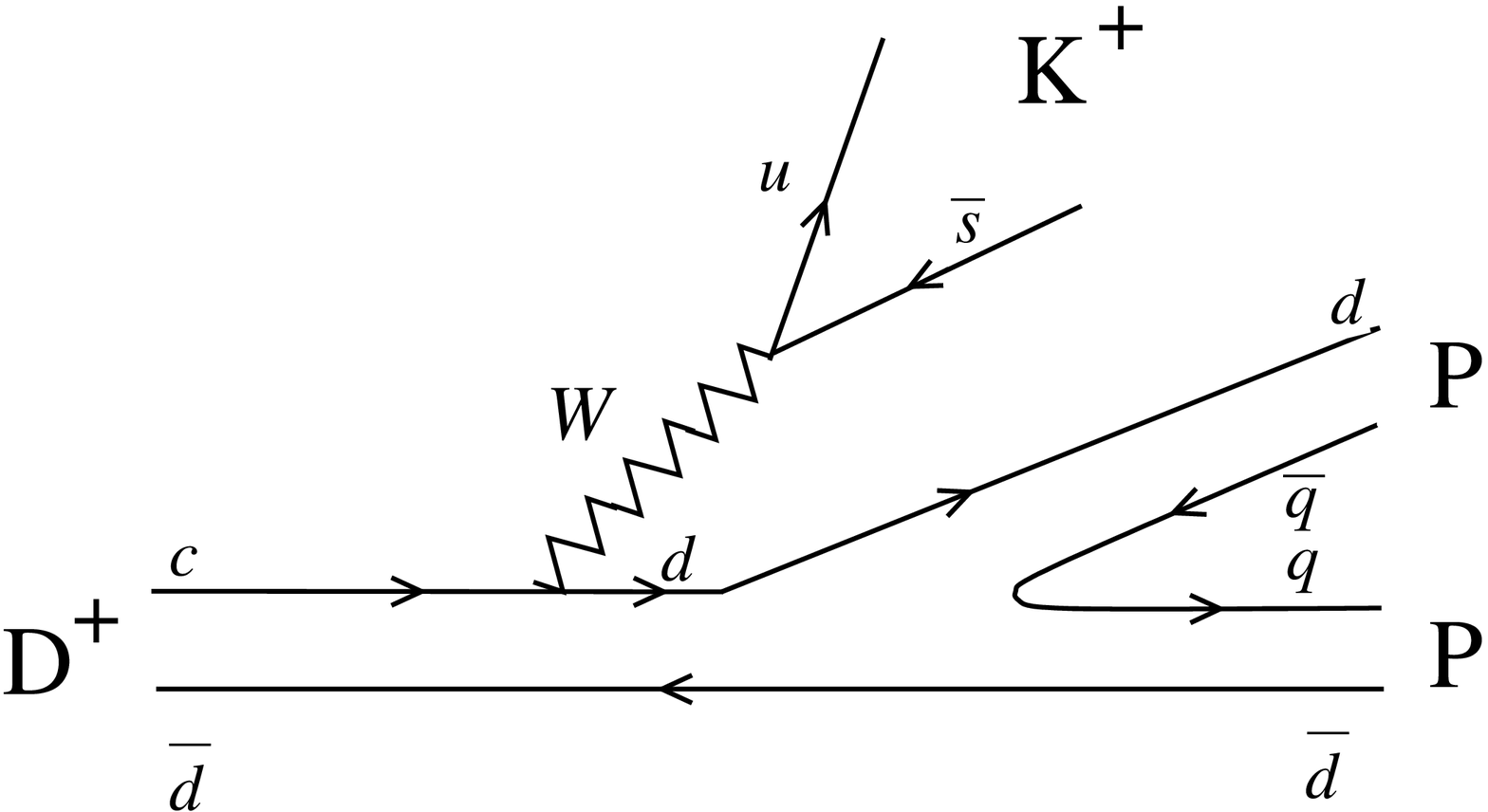}} \\
     \subfigure[]{\includegraphics[width=.8\linewidth]{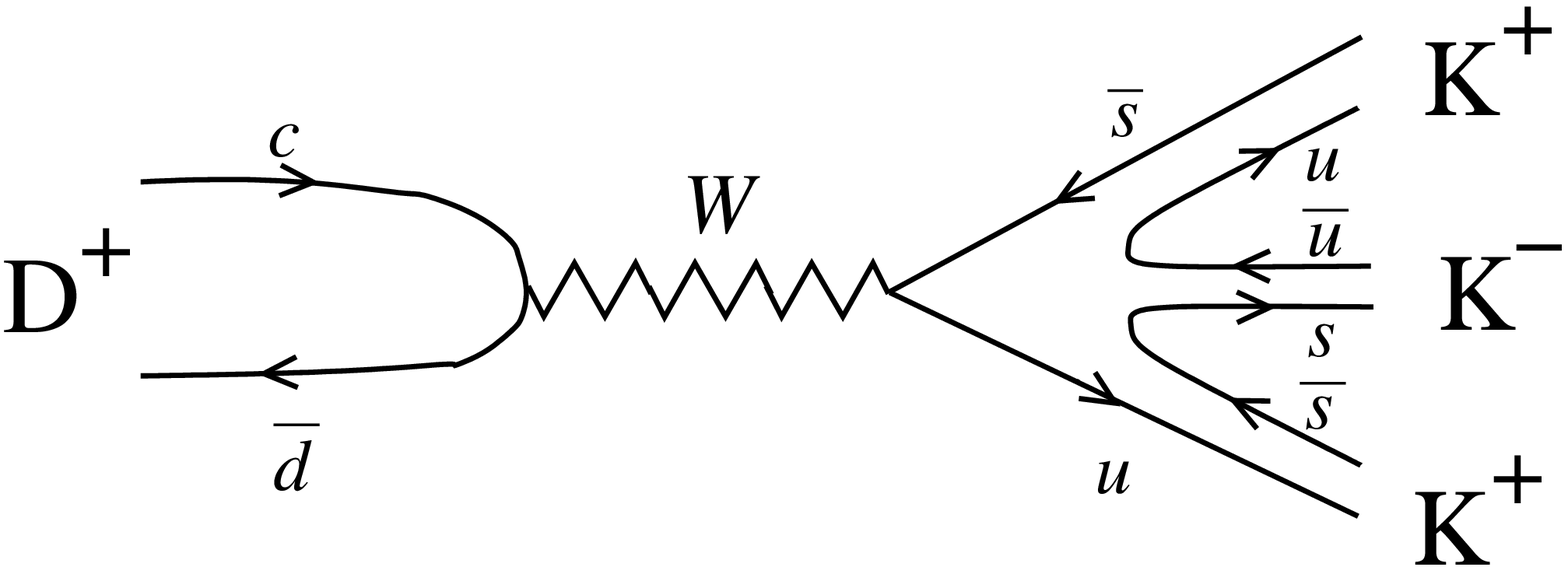}} 
    \caption{\small{Elementary $D^+\to K^+ P P$ process at the quark level.}}
\label{fig:Tquarks}
\end{figure}

As briefly discussed in the Introduction,
the possible elementary quark topologies at tree level for the  $D^+\to K^- PP$ process, where $PP$ stands for a pseudoscalar meson pair, are depicted in Fig.~\ref{fig:Tquarks}.
Note that $PP$ in Fig.~\ref{fig:Tquarks}(a) cannot be $K^+K^-$ at the tree level since it cannot be produced from $\bar d d$, but  the $K^+K^-$ pair can be produced via final  state interaction (FSI) of the pseudoscalar pair. This necessity for FSI was the main reason for neglecting the diagram (a) in Ref.~\cite{Aoude:2018zty}, which is the basis of the LHCb experimental analysis in
Ref.~\cite{Aaij:2019lwx}. However, our position in the present work is opposite and we are going to argue why we expect the (a) diagram to be dominant. First, diagram (b) represents annihilation and, 
since the $D$ have spin zero, the $W$
annihilation diagram is suppressed by helicity conservation at the light
quark vertex
 \cite{Morrison:1989xq}. In addition, diagram (b) requires two hadronizations, each of which reduces the width by about one order of magnitude \cite{Bayar:2014qha}.
Furthermore, diagram (a) relies upon external emission, which has the largest strength for weak interaction \cite{Morrison:1989xq}. Indeed, one quark is operative and the other one remains spectator which implies a one body operator, versus the two body operator required in the annihilation, and is color favored. On the other hand, the FSI interaction necessary to produce final $K^+K^-$ is actually required at low invariant masses since that region is much influenced by the $a_0(980)$ and $f_0(980)$ resonances which are dynamically generated within the chiral unitary approach (UChPT) from the final $PP$ interaction, as explained in the Introduction.
We will come back to the implementation of the FSI trough the UChPT amplitudes in the second part of this section, but first we address the calculation of the 
elementary production at the quark level in Fig.~\ref{fig:Tquarks}(a).

While the $\bar d$ quark remains as spectator, the $c$ quark becomes a $d$  through the emission of a $W$ boson, which eventually creates the $u\bar s$ of a $K^+$. Note that this process involves two Cabibbo suppressed weak transitions, ($Wcd$, $Wus$).  
The final $\bar d d$ pair then hadronizes into a final pseudoscalar meson pair, which is implemented by producing an extra $\bar{q} q$ with  the $^3P_0$ model  \cite{micu,LeYaouanc:1972vsx,bijker}.
The weight of the different allowed pseudoscalar pairs produced in the hadronization, can be related, up to a global normalization factor, using the following $SU(3)$ arguments:

Let $|H\rangle$ be the flavor state  of the final hadronic part after the quark-antiquark pair is produced in the hadronization:
\begin{align}
|H\rangle \equiv |d\,(\bar u u +\bar d d +\bar s s)\,\bar d \rangle.
\label{eq:uuddss}
\end{align}
It can be written as 
\begin{align}
|H\rangle = \sum_{i=1}^3{|d \bar q_iq_i\bar d} \rangle=
            \sum_{i=1}^3{|M_{2i}M_{i2} }    \rangle=
	     |(M^2)_{22}\rangle,
\label{eq:Hket}
\end{align}
where we have defined

\begin{equation}
q\equiv \left(\begin{array}{c}u\\d\\s\end{array}\right)\,\text{~~and~~~}
M\equiv q\bar q^\intercal=\left(\begin{array}{ccc}u\bar u & u\bar d & u\bar s\\
					       d\bar u & d\bar d & d\bar s\\
					       s\bar u & s\bar d & s\bar s
\end{array}\right)\,.
\label{eq:Mqqbar}
\end{equation} 

The strength of $SU(3)$ comes into play when we associate the matrix $M$ to the usual $SU(3)$ matrix containing the pseudoscalar mesons:

\begin{align*}
M\Rightarrow P\equiv
\left(\begin{array}{ccc} 
              \frac{\pi^0}{\sqrt{2}}  + \frac{\eta}{\sqrt{3}}+\frac{\eta'}{\sqrt{6}}& \pi^+ & K^+\\
              \pi^-& -\frac{1}{\sqrt{2}}\pi^0 + \frac{\eta}{\sqrt{3}}+ \frac{\eta'}{\sqrt{6}}& K^0\\
              K^-& \bar{K}^0 & -\frac{\eta}{\sqrt{3}}+ \frac{2\eta'}{\sqrt{6}} 
      \end{array}
\right)\,,
\label{eq:Pmatrix}
\end{align*}
where we have used ideal mixing between the singlet and octet to 
give $\eta$ and $\eta'$ \cite{Bramon:1992kr}. 
Then, the matrix element required in Eq.~\eqref{eq:Hket} is 

\begin{equation}
(P^2)_{22}=
\pi^- \pi^+ +   \frac{1}{2} \pi^0 \pi^0
+  \frac{1}{3}\eta\eta-\sqrt{\frac{2}{3}}\pi^0\eta +K^0 \bar{K}^0.
\label{eq:weightsPP}
\end{equation}

\begin{figure}[h]
\begin{center}
\includegraphics[width=0.5\textwidth]{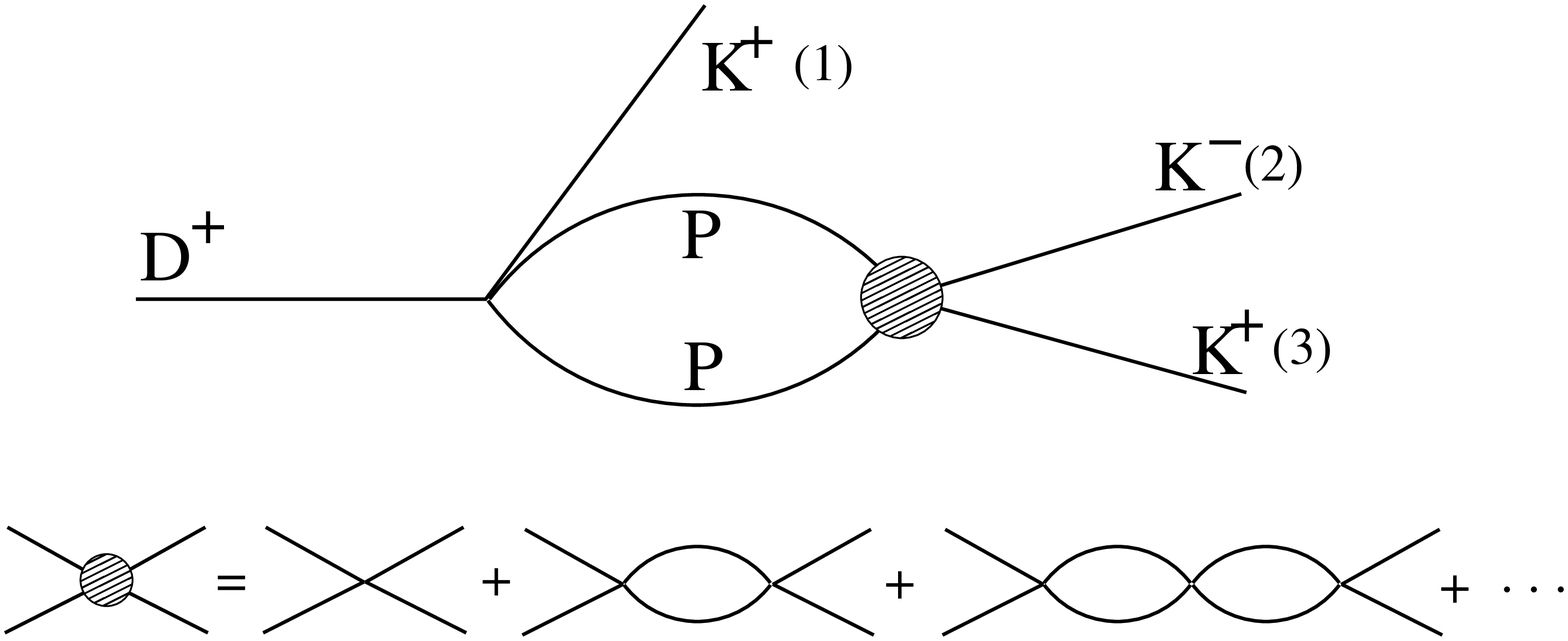}
\caption{\small{Final state interaction of the $D^+\to K^+ P P$ process to get $D^+\to K^- K^+K^+$.}}
\label{fig:TFSI}
\end{center}
\end{figure}

Note that, as mentioned above, no $K^+K^-$ pair is possible in the  hadronization 
from $\bar d d$
and then it must necessarily be produced in the final state interaction from the five possible pseudoscalar pairs,
$\pi^- \pi^+$, $\pi^0 \pi^0$, $\eta\eta$, $\pi^0\eta$ and $K^0 \bar{K}^0$,
 as depicted in Fig.~\ref{fig:TFSI}.

Note that we are not including the scalar $f_0(980)$ and $a_0(980)$   resonances as explicit degrees of freedom but they arise naturally in the non-linear dynamics involved when implementing unitarity in coupled channels starting from a lowest order tree level meson-meson chiral potential. This effectively accounts for the resummation shown in Fig.~\ref{fig:TFSI} and it is the basis of the chiral unitary approach (UChPT). In the scalar sector, there are several different ways to implement these ideas like the Bethe-Salpeter equation \cite{Oller:1997ti}, the inverse amplitude method \cite{Oller:1998hw,dobado-pelaez} or the N/D method \cite{Oller:1998zr}, but all of them provide similar results. 
Since in the present work we are going to compare with experimental data from the LHCb collaboration which involves $K^- K^+$ invariant masses up to 1375$\mev$, we  use the amplitudes from the N/D approach 
\cite{Oller:1998zr} which provides the largest range of predictability among the aforementioned approaches. 
All theses approaches rely upon mainly one free parameter coming from the regularization, either a cutoff or a subtraction constant, which is determined from a fit to meson-meson scattering data. The N/D method of  \cite{Oller:1998zr}  is able to extend the applicability range up to higher energies by including in the interaction kernel, in addition to the lowest order ChPT
amplitudes, the s-channel exchange of scalar resonances in a chiral symmetric invariant way. These tree level resonances constitute an octet with mass around $1.4\gev$ and a singlet
around $1\gev$ which barely changes the dynamical origin of the $a_0(980)$ and $f_0(980)$ but improves the amplitude close to 1400$\mev$.
In any case, all of the UChPT approaches provide similar results in the region around 1~GeV.
\begin{figure}[h]
\begin{center}
\includegraphics[width=0.5\textwidth]{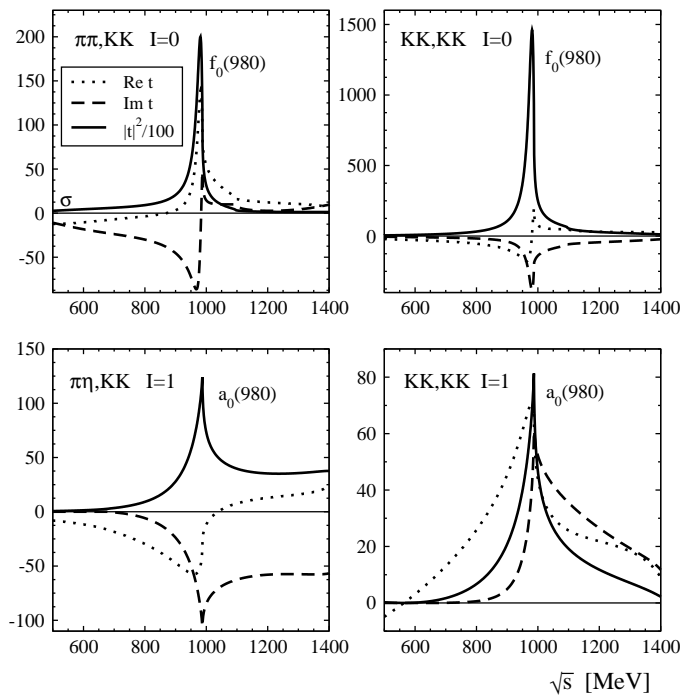}
\caption{\small{Meson-meson scattering amplitudes for isospin $I=0$ and $I=1$.}}
\label{fig:Tmmmm}
\end{center}
\end{figure}
In Fig.~\ref{fig:Tmmmm} we show some spin 0 and isospin $I=0$ and $I=1$ meson-meson scattering amplitudes\footnote{
 In Fig.~1 of Ref.~\cite{Oller:2006xb} a comparison with several scattering data obtained using these amplitudes can be seen.}, which will be needed in the present work. The energy range involved in the present work is from the $K^+K^-$ threshold, $987 \mev$, till $M_D-m_K= 1375\mev$. We can clearly see the shapes for the  $f_0(980)$ and $a_0(980)$, but note that these shapes are  far from being just Breit-Wigners, and this is one of the strong points of UChPT: it provides not only the pole structure of the resonances but the actual scattering amplitude. The  $a_0(980)$ actually corresponds to a cusp at the $K\bar K$ threshold.

We can then write the amplitude corresponding to the process in Fig.~\ref{fig:TFSI}. If we use the label 1 for the $K^+$ coming directly from the $D^+$, label 2 for the $K^-$ and 3 for the other $K^+$ (see Fig.~\ref{fig:TFSI}), it can be written as 
\begin{equation}
T(s_{23})={\cal C} \sum_{i=1}^5 h_i\, G_i(s_{23})\, t_{i,K^+K^-}(s_{23})
\label{eq:T1}
\end{equation}
where  $s_{ij}=(p_i+p_j)^2$. In Eq.~\eqref{eq:T1} 
the sum runs over the five $PP$ allowed channels in Eq.~\eqref{eq:weightsPP}, 
${\cal C}$ is an arbitrary global normalization factor to be fitted  later on to the experimental LHCb data, 
 $h_i$ are the numerical coefficients in front of each $PP$ channel in Eq.~\eqref{eq:weightsPP}, $t_{i,K^+K^-}$ stands for the unitarized $(PP)_i\to K^+K^-$ amplitude in s-wave explained above and $G_i$ is the loop function for two pseudoscalar mesons regularized with the same subtraction constant used in the evaluation of  $t_{i,K^+K^-}$.
 
We can theoretically filter the different isospin contributions taking into account that $\pi^- \pi^+$, $\pi^0 \pi^0$ and  $\eta\eta$ contribute only to $I=0$, and $\pi^0\eta$ to $I=1$. The $K^0 \bar{K}^0$ pair contributes to both isospins, but taking into account the isospin decomposition of the different $K \bar K$ states
\begin{align}
| K^+K^- \rangle&= -\frac{1}{\sqrt{2}}|K\bar K\rangle_{I=1, I_3=0} -\frac{1}{\sqrt{2}}|K\bar K\rangle_{I=0, I_3=0},\nn\\
| K^0 \bar K^0 \rangle&= \frac{1}{\sqrt{2}}|K\bar K\rangle_{I=1, I_3=0} -\frac{1}{\sqrt{2}}|K\bar K\rangle_{I=0, I_3=0},
\label{eq:changebasis}
\end{align}
it suffices to substitute, in  Eq.~\eqref{eq:T1},
\begin{align} t_{K^0 \bar K^0,K^+K^-}\to \frac{1}{2}(t_{K^0 \bar K^0,K^+K^-}+t_{K^+K^-,K^+K^-})
\end{align}
 for $I=0$, and
\begin{align} t_{K^0 \bar K^0,K^+K^-}\to \frac{1}{2}(t_{K^0 \bar K^0,K^+K^-}-t_{K^+K^-,K^+K^-})
\end{align}
for $I=1$.

 To the $I=0$ amplitude in Eq.~\eqref{eq:T1} we must add the contribution from the $\phi(1020)$ meson which, being a genuine $\bar q q $ resonance in p-wave, is no included in the aforementioned meson-meson scattering amplitudes. Since the amplitude is going to span a large invariant mass region, we consider a full relativistic amplitude as in Ref.~\cite{Toledo:2020zxj}:
 
 \begin{equation}
T_\phi(s_{23},s_{12})={\cal D}\frac{s_{13}-s_{12}}{s_{23}-m_\phi^2+im_\phi\Gamma_\phi(\sqrt{s_{23}})}
\label{eq:Tphi}
\end{equation}
where $s_{13}=
M_{D}^2+3m_K^2-s_{12}-s_{23}$,
$D$ is a complex arbitrary normalization factor to be fitted later on, and we use an energy dependent p-wave $\phi$ width 
 \begin{equation}
 \Gamma_\phi(m)=\Gamma_o \frac{m_\phi}{m}\frac{p^3(m)}{p^3(m_\phi)}B^2(m)
\end{equation}
 where $\Gamma_o$ is the total width of the $\phi$, $p(m)$ the $K$  momentum in the $\phi$ rest frame for a $\phi$ invariant mass $m$,   and  $B(m)$ is the p-wave Blatt-Weisskopf barrier penetration  factor \cite{blattweisskopf} given by
\be
B(m)=\left(\frac{1+(R\, p(m_\phi))^2}{1+(R\, p(m))^2}\right)^{1/2}
\label{eq:blatt}
\ee
In Eq.~\eqref{eq:blatt}, $R$ stands for the range parameter of the $\phi$ a typical value of $R=1.5\,\textrm{GeV}^{-1}$, although it is not very relevant.
 
Finally, the three body distribution for the $D^+\to K^- K^+K^+$ decay is given by
 \begin{equation}
 \frac{d^2\Gamma}{d{s_{12}}d{s_{23}}}=
 \frac{1}{32(2\pi)^3M_{D}^3}\frac{1}{2}|{\cal{M}}|^2
 \label{eq:dalitzM}
 \end{equation}
 where ${\cal{M}}$ is the total $D^+\to K^- K^+K^+$ amplitude considered adding Eqs.~\eqref{eq:T1} and \eqref{eq:Tphi} which must be properly symmetrized (exchanging the labels $1\longleftrightarrow 3$) since we have two identical $K^+$ in the final state:
  \begin{equation}
 {\cal{M}}(s_{23},s_{12})=T(s_{23})+T_\phi(s_{23},s_{12})+(1\longleftrightarrow 3).
\end{equation}
from where the $K^+ K^-$ spectrum can be obtained as
\begin{align}
\frac{d\Gamma}{ds_{K^+ K^-}}=\int_{s_{23}^\textrm{min}}^{s_{23}^\textrm{max}}
d{s_{23}}
\frac{d^2\Gamma}{d{s_{12}}d{s_{23}}}.
\label{eq:dGsKK}
\end{align}

In the results section we will also evaluate the $K^+K^+$ distribution which is given by
\begin{align}
\frac{d\Gamma}{ds_{K^+K^+}}=\int_{s_{23}^\textrm{min}}^{s_{23}^\textrm{max}}
d{s_{23}}
\frac{d^2\Gamma}{d{s_{13}}d{s_{23}}},
\end{align}
but using for $s_{12}$ in the argument of ${\cal{M}}(s_{23},s_{12})$, 
$s_{12}=
M_{D}^2+3m_K^2-s_{13}-s_{23}$.

\section{Results}

\begin{figure*}[h]
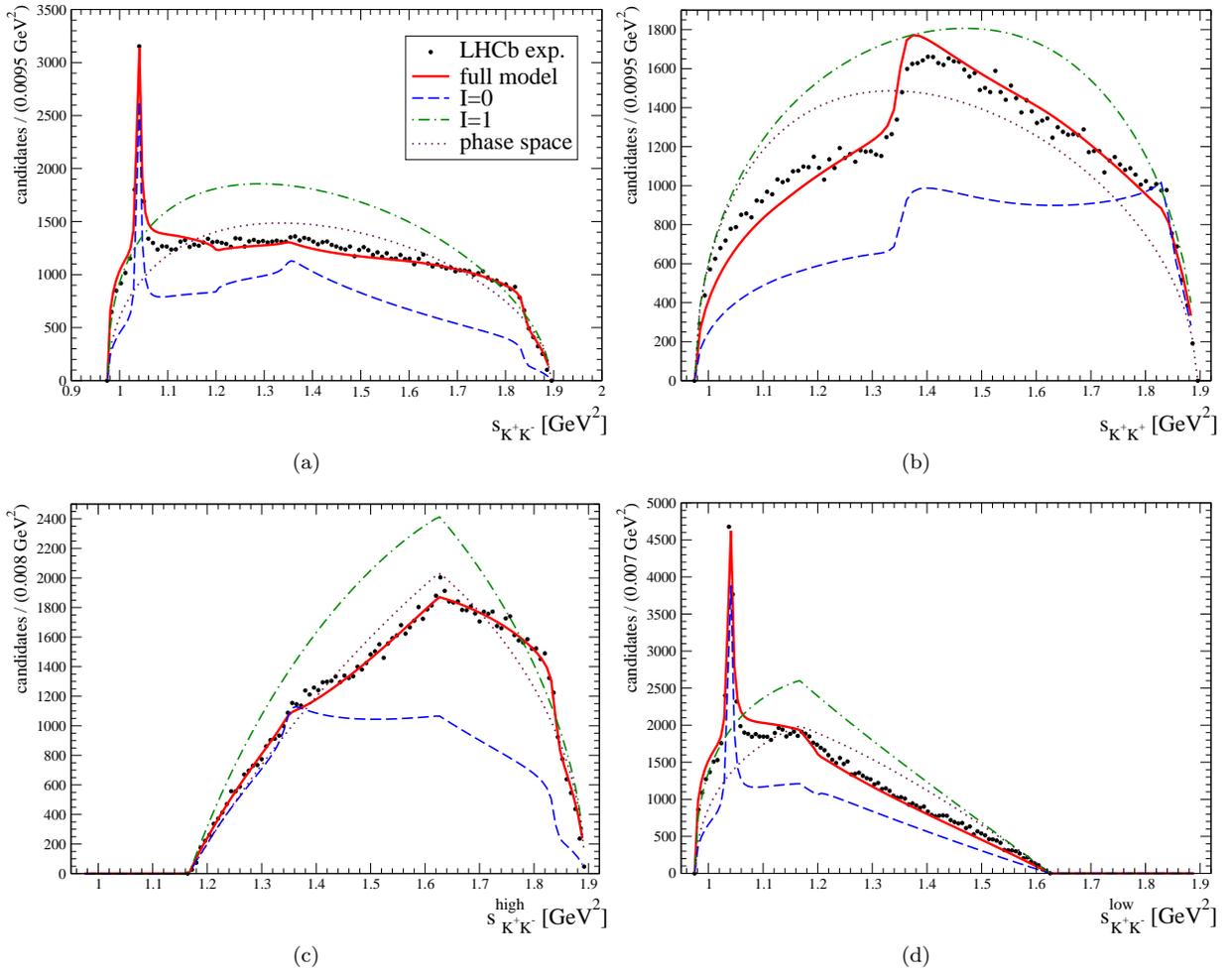

     \centering
     \subfigure[]{\includegraphics[width=.45\linewidth]{resultKMKm.eps}}
     \subfigure[]{\includegraphics[width=.45\linewidth]{resultKpKp.eps}} \\
     \subfigure[]{\includegraphics[width=.45\linewidth]{resultKMKm_high.eps}}
     \subfigure[]{\includegraphics[width=.45\linewidth]{resultKMKm_low.eps}} \\
   \caption{\small{ (Color online) Invariant mass distributions  in comparison with the experimental data from Ref.~\cite{Aaij:2019lwx}
}}
\label{fig:results1}
\end{figure*}

\begin{figure}[h]
\begin{center}
\includegraphics[width=0.4\textwidth]{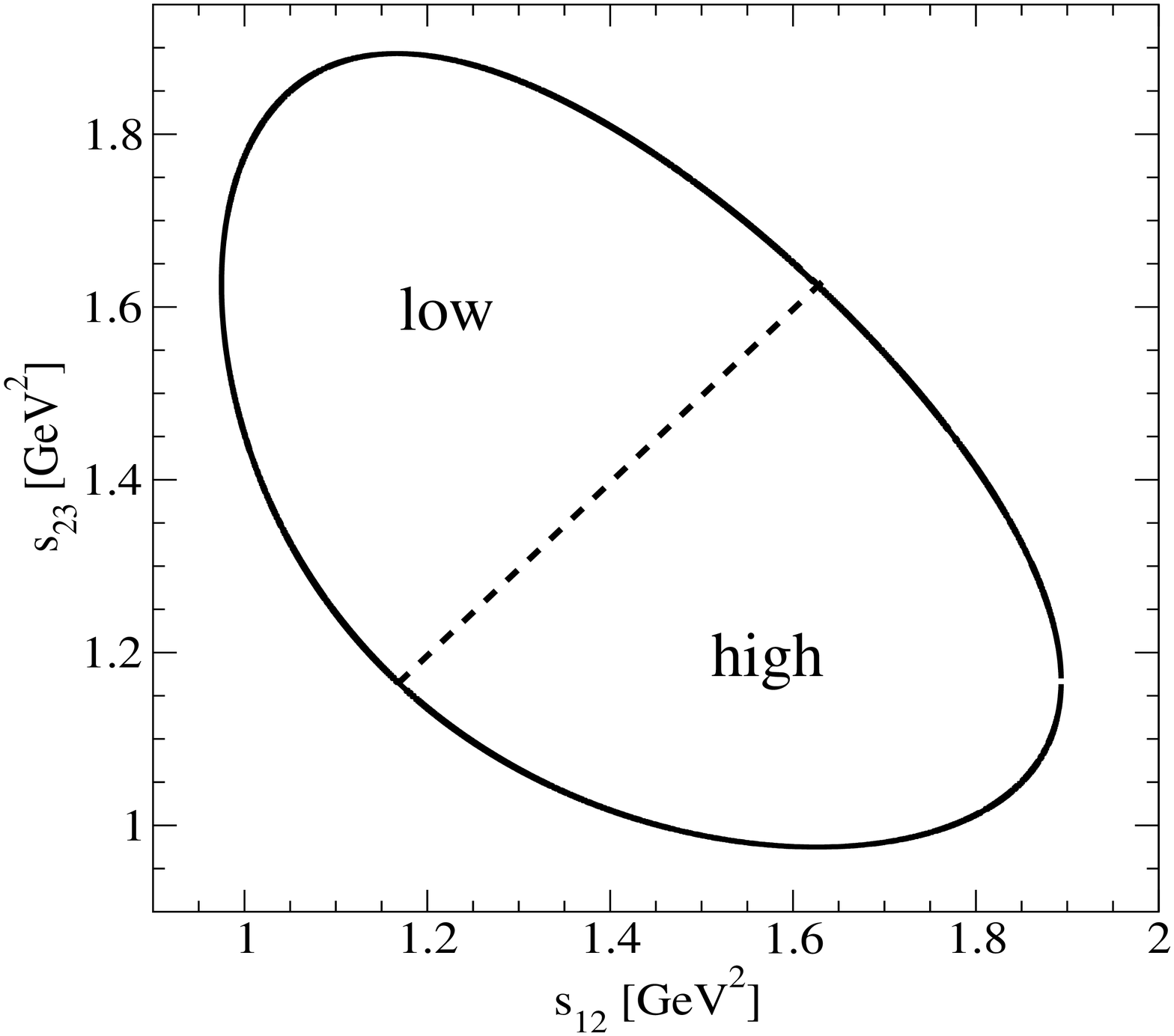}
\caption{\small{Dalitz plot with the definition of the regions $s_{12}^\textrm{high}$ and $s_{12}^\textrm{low}$.}}
\label{fig:dalitz}
\end{center}
\end{figure}

Our model has three parameters: one for the global normalization
 ${\cal C}$ in Eq.~\eqref{eq:T1} and two for the global weight of the $\phi$ meson amplitude, complex ${\cal D}$ in Eq.~\eqref{eq:Tphi}, which we fit to the experimental \cite{Aaij:2019lwx} $K^+K^-$ distribution, (only panel-(a) in Fig.~\ref{fig:results1}). The other panels in Fig.~\ref{fig:results1} represent the $K^+K^+$ distribution and the distributions $s_{K^+K^-}^\textrm{high}$ and $s_{K^+K^-}^\textrm{low}$ where, 
according to  \cite{Aaij:2019lwx}, $s_{K^+K^-}^\textrm{high}$ and $s_{K^+K^-}^\textrm{low}$ represent the highest and lowest values among $s_{12}$ and $s_{23}$, see Fig.~\ref{fig:dalitz}.  Theoretically, we evaluate the $s_{K^+K^-}^\textrm{low}$ distribution including $\theta(s_{23}-s_{12})$ in the integrand of  Eq.~\eqref{eq:dGsKK}, with $\theta$ the step function.

In all the figures the dots represent the LHCb experimental data 
\cite{Aaij:2019lwx}, the solid line our full model, the dotted line the phase-space, the dashed line the $I=0$ contribution and the dashed-dotted line the $I=1$.
The experimental data in  Ref.~\cite{Aaij:2019lwx} are not corrected for setup acceptance, however phase-space curves  weighted by the efficiency for the different plots in Fig.~\ref{fig:results1} are provided in the experimental paper. Therefore, we have renormalized each experimental datum  such that the phase-space agrees with the theoretical three body  distribution.

A first observation from Fig.~\ref{fig:results1} is that our model fits reasonably well
the whole spectrum of the $ K^+K^-$, $ K^+K^+$  distributions (recall that we have only fitted the data of panel a). This is remarkable, given the little freedom in the fit: just the global normalization factor and the relative complex weight of the $\phi(1020)$. The rest is given from the non-trivial unitarization model  implied in  Eq.~\eqref{eq:T1}. It is also worth recalling again that there is no $ K^+K^-$ in the elementary production vertex of Fig.~\ref{fig:Tquarks}(a) and thus all the strength is coming from the FSI starting with meson-meson channels other than $ K^+K^-$.

On the other hand, by looking at the different isospin contributions in Fig.~\ref{fig:results1},  we see that the $I=1$ contribution dominates over the $I=0$ one. In particular, close to threshold
the accumulation of the strength with respect to the phases-space is mainly due to the $I=1$ amplitude, {\it i.e.} the effect of the $a_0(980)$.
\begin{figure}[h]
\begin{center}
\includegraphics[width=0.5\textwidth]{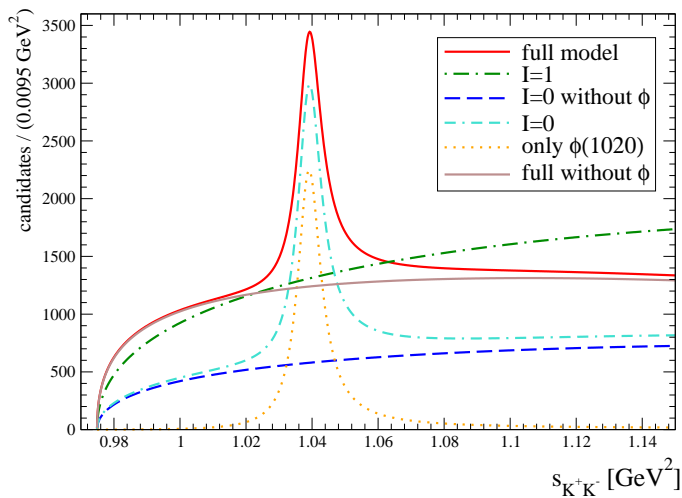}
\caption{\small{(Color online) Different contributions to the $D^+\to K^- K^+K^+$ mass distribution close to threshold.}}
\label{fig:zoom}
\end{center}
\end{figure}
This is more clearly manifest if we look at Fig.~\ref{fig:zoom}, where we zoom in the $K^+K^-$ mass distribution  near the threshold, including, in addition to the theoretical curves of  Fig.~\ref{fig:results1}(a), the contribution considering only the $\phi$ meson, only the $I=0$ without the $\phi$, and the full model removing the $\phi$.
This is also more clearly seen if we theoretically remove the phase-space by plotting, in Fig.~\ref{fig:M},
the different isospin contributions from the FSI terms, {\it i.e.} without the $\phi$, to the squared
amplitude $|{\cal M}|^2$ of Eq.~\eqref{eq:dalitzM} without the $1\longleftrightarrow 3$ symmetrization, $i.e$, ${\cal M}\equiv 
T(s_{23})$, as a function of $s_{23}$. Note that if we included the $\phi$ meson or the symmetrization, the amplitude would also depend on the $s_{12}$ variable. We see that close to threshold, indicated by the vertical dotted line, the strength is essentially dominated by the $a_0$ contribution. Below threshold the shapes of the $a_0$ and $f_0$ are clearly visible in this plot but are not accessible when considering the actual phase-space.
\begin{figure}[h]
\begin{center}
\includegraphics[width=0.5\textwidth]{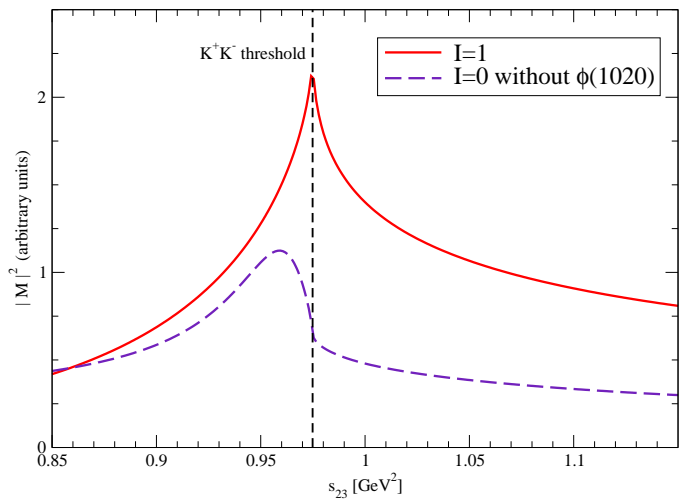}
\caption{\small{(Color online) Different isospin contributions to the $D^+\to K^- K^+K^+$ amplitude. The vertical dashed line represents the $ K^+K^-$ threshold.}}
\label{fig:M}
\end{center}
\end{figure}
In Fig.~\ref{fig:zoom}, at low invariant masses we see a pattern for the $f_0(980)$ and $a_0(980)$ contributions different than what was found in Ref. \cite{Aoude:2018zty}. In both approaches a dominance of the $a_0(980)$ contribution is found. In Ref.~\cite{Aoude:2018zty} a destructive interference between the $a_0$ and $f_0$ contributions was reported. We find  a different pattern. The addition of the $f_0(980)$ resonance increases  the contribution of the $a_0(980)$ at low invariant masses but subtracts for invariant masses higher than 1020~MeV. The interference of the $I=0$ and $I=1$ contributions is made possible in the present work because both contributions appear in the variables $s_{12}$ and $s_{23}$.

\section{Summary}
We show theoretically that the $D^+\to K^- K^+K^+$ decay can be understood from the mechanism that accounts for the final state interaction of an initial pseudoscalar pair (see Fig.~\ref{fig:Tquarks}(a)). Indeed  $K^- K^+K^+$ in the final state is not possible at the tree level and hence the rescattering is mandatory. At this point we take advantage of the unitary extensions of chiral perturbation theory which generate dynamically the $f_0(980)$ and $a_0(980)$ resonances, without the need to include them as explicit degrees of freedom, and provide the full meson-meson scattering amplitudes, not only the resonances. The relative weights of the initial production of the meson-meson pairs are obtained from SU(3) arguments, and then  for the unitarization we use the UChPT amplitudes from the N/D method which allows to extend the range of applicability to the whole final $K^+K^-$ mass spectrum  in this decay. With a minimal freedom, just the global normalization and the weight and phase of the $\phi(1020)$ contribution, we are able to fit experimental data on $K^+K^-$ invariant mass distribution.
 We also show that the dominant contribution close to threshold comes from the $I=1$ (hence the $a_0(980)$) contribution in clear dominance over the $I=0$ one (the $f_0(980))$.
 The remarkable agreement is a step in favor of considering this mechanism as the leading one in this decay, at odds with other considerations as in the experimental analysis \cite{Aaij:2019lwx} based on the work in Ref.~\cite{Aoude:2018zty} which advocates for the dominance of the mechanism in Fig.~\ref{fig:Tquarks}(b), and to reinforce the dynamical origin of the $a_0(980)$ and $f_0(980)$.

\section{Acknowledgments}
One of us, E. O., wishes to acknowledge useful discussions with Alberto Correa dos Reis and his encouragement for us to study this reaction. 
 This work is partly supported by the National Natural Science Foundation of China under Grant Nos. 11505158, 11847217, 11975083 and 11947413.  It is also supported by the Academic Improvement Project of Zhengzhou University.
This work is partly supported by the Spanish Ministerio
de Economia y Competitividad and European FEDER funds under Contracts No. FIS2017-84038-C2-1-P B
and No. FIS2017-84038-C2-2-P B.
This project has received funding from the European Union's Horizon 2020 research and innovation programme under grant agreement No 824093 for the STRONG-2020 project.

\end{document}